\documentclass[12pt,reqno]{amsart}
\usepackage{graphicx}
\usepackage{subfigure}
\usepackage[T1]{fontenc}
\usepackage{enumerate}
\usepackage{amsmath}
\usepackage{amsfonts}
\usepackage{amssymb}
\usepackage[colorlinks=true, linkcolor=blue, citecolor=blue, urlcolor=blue]{hyperref}
\usepackage{placeins}
\usepackage{scrextend}
\usepackage{fancyhdr}
\usepackage{xcolor}
\usepackage[utf8]{inputenc}
\allowdisplaybreaks
\makeatletter
\def\section{\@startsection{section}{1}%
	\z@{.7\linespacing\@plus\linespacing}{.5\linespacing}%
	{\normalfont\bfseries}}
\def\subsection{\@startsection{subsection}{2}%
	\z@{.5\linespacing\@plus.7\linespacing}{-.5em}%
	{\normalfont\bfseries}}
\makeatother

\usepackage{caption}
\usepackage{morefloats}
\newtheorem{thm}{Theorem}[section]

\numberwithin{equation}{section}

\title[]{Optimal systems, conservation laws, and invariance analysis of the $(2+1)$ extended Boiti-Leon-Manna-Pempinelli equation via the Lie symmetry approach}
\author[]{}
\begin{document}
	\leftline{ \scriptsize \it }
	\maketitle
	\begin{center}
		\textbf{Akshita Bhardwaj}\footnote{\textbf{Email:} akshita\_b@as.iitr.ac.in}, \textbf{Shalini Yadav}\footnote{\textbf{Email:} shalini.yadav@lsr.du.ac.in}, \textbf{Muhammad Junaid-U-Rehman}\footnote{\textbf{Email:} muhammad-junaid.u-rehman@dokt.p.lodz.pl} and \textbf{Rajan
			Arora}\footnote{ \textbf{Email:} rajan.arora@as.iitr.ac.in,
			\textbf{Corresponding author}}\\
		
		\textsuperscript{1,4}Department of Applied Mathematics and Scientific Computing,\\
		Indian Institute of Technology
		Roorkee, India\\
        \textsuperscript{2}Department of Mathematics,\\
		Lady Shri Ram College For Women–Delhi University\\
		\textsuperscript{3}Department of Automation, Biomechanics, and Mechatronics,\\ Lodz University of Technology, Lodz, Poland
	\end{center}
\begin{abstract}
			Lie symmetry analysis has been applied to the extended Boiti-Leon-Manna-Pempinelli (eBLMP) equation. This system illustrates the exchange of information between two waves with distinct dispersion characteristics. The optimal system of the corresponding Lie algebra has been constructed. The equation considered has been reduced into a simpler form for the computation of analytical solutions. The novelty of this research is the optimal system of subalgebras in one dimension using the adjoint action approach. To analyze and understand the eBLMP more clearly, graphs have been plotted. We have also found conservation laws.
	\end{abstract}

	\section{Introduction}
	Sophus Lie was an accomplished scientist in the 19th century. His peers described him as an exceptionally gifted mathematician with a penetrating mind and creative imagination. Lie preferred the geometric
	approach to mathematical problems over the arithmetic approach, which
	is quite visible in his work.\hspace{0.1cm}Lie was most intrigued by group theory
	and its applications in differential equations.\hspace{0.1cm}His research on
	inline-sphere transformation and the formulation of the theory of
	continuous groups and their application to various mathematical areas
	was innovative and consequently led to the discipline of mathematics
	known as Lie theory \cite{1}.\hspace{0.1cm}In the literature review of Lie theory, we observed that a few publications that discuss group invariant solutions of partial differential
	equations (PDEs) had been published in over three decades (\cite{2,3,4,30}). There are several advantages to studying nonlinear waves, specifically equations such as the extended Boiti-Leon-Mana-Pempinelli (BLMP) equation, which falls under the category of shallow water wave equations. Linear wave equations frequently do not adequately represent significant characteristics of wave systems in the actual world, such as abrupt changes in gradient, the creation of shocks, and the dispersion of effects. Nonlinear wave equations are crucial for realistic modeling as they offer more precise descriptions of these processes. Nonlinear wave equations frequently display intricate phenomena, including the production of solitons, wave breaking, and wave interactions. These phenomena have practical ramifications and can aid in comprehending the transfer and dissipation of energy in wave systems. Applications in the field of engineering: Nonlinear wave equations find application in diverse engineering fields, including the design of wave energy converters, the optimization of coastal structures, and the prediction of wave impacts on offshore structures. Recent research in PDEs has led to interesting work (see \cite{301,302,303}), and applications in fractional equations as well (\cite{31},\cite{32},\cite{33}). Solitons, a particular type of exact solution, have also gained quite a bit of popularity. These solutions help understand the behavior of light pulses in optical fibers and non-linear and dispersive phenomena in water wave systems (\cite{42,43,44,45,46}). 
	In \cite{51}, Bluman and Anco provided a new area of research for
	researchers in the fields of Lie theory and conservation laws. In
	\cite{15}, Olver gave a lucid description of
	the geometrical aspect of Lie theory. The Lie symmetry technique is demonstrated in the paper \cite{8} for a particular class of PDEs called the evolution equation.\hspace{0.1cm}An evolution equation is a PDE that describes how a system evolves with time using information from some given
	initial data.\hspace{0.1cm}Evolution equations have applications in several areas of applied and engineering sciences.\hspace{0.1cm}A particular example of a non-linear evolution
	equation is the Ablowitz-Kaup-Newell-Segur water wave equation
	(AKNSWW). In physical science, the AKNSWW dynamical equation is relevant. The generalized bilinear approach of the AKNSWW equation \cite{9} gives rise to the Bolti-Leon-Manna-Pempinelli (BLMP) equation.\hspace{0.1cm}In 
	\cite{17}, the $(2+1)$-dimensional Boiti-Leon-Manna-Pempinelli (BLMP) equation has been implemented in an alternative form of the Nizhnik-Novikov-Veselov (NNV) equation and Lie transformation group theory.\hspace{0.1cm}The results of this research article
	entail some exact but generalized solutions to $(2+1)$-dimensional Boiti-Leon-Manna-Pempinelli through the theory of Lie transformation groups.\hspace{0.1cm}There is a window for improvements in every research work, and here we are considering, in particular, exact
	solutions and symmetries.\hspace{0.1cm}Because of the complex nature of the motion of shallow water
	waves, dealing with them requires a significant amount of effort. This complex nature resonates with most of the phenomena in
	nature, which makes it important.\\
   \vspace{-0.09mm}
  The limitations of this work are equally important to consider as well. Symmetry approaches and optimal systems may not offer a comprehensive understanding of the behavior of complicated systems. Although symmetry approaches are useful in detecting symmetries and invariant solutions of PDEs, they may not always encompass the complete spectrum of behaviors displayed by nonlinear equations such as the eBLMP problem. Additionally, it is difficult to construct lie algebra for every non-linear system.\\
  \vspace{-0.09mm}
	Olver and Ovsiannnikov \cite{15,18} put forth the concept of an
	optimal system for Lie symmetry analysis.\hspace{0.1cm}Patera et al.\hspace{0.1cm}\cite{20,21}
	did remarkable work in this area and paved the way for other authors
	to find optimal systems for several systems. The optimal system of the eBLMP equation has been constructed in this paper
	using the method elaborated in \cite{22, 23, 24, 11125}. 
	\subsection{Research relevant to the eBLMP equation}
The eBLMP equation was first introduced in \cite{101}, which also
	demonstrates the application of Lie symmetries for the analysis of the
	group properties for the $(2+1)$ eBLMP equation. New periodic solutions are also determined.\hspace{0.1cm}In \cite{102}, a class of lump and kink solutions of the system \eqref{eq:1} has been put forward.\hspace{0.1cm}In\hspace{0.1cm}\cite{103}, standard
	Weiss Tabor Carnevale's approach and Kruskal's simplification have
	been used to justify the painlevé non-integrability of the equation.\\
  \vspace{-0.09mm}	
 The BLMP equation is used to define non-linear wave properties in shallow water wave equations \cite{9,111,112,113,114}. More precisely, it outlines the interactions of waves having
	different dispersion relations, this physical application makes the equation quite relevant. The $(2+1)$-dimensional BLMP equation, a modified version of the shallow-water wave equation, was developed using the bilinear strategy. B\"acklund transformations are one of the BLMP equation's intriguing characteristics \cite{117}. The exact solutions of the BLMP equation and other physical
	applications can be found in \cite{113, 114, 121}, while exact
	solutions with the method of Lie symmetries were determined in \cite{17}. The $(2+1)$ eBLMP was proposed in \cite{101}.
	\begin{align}
		\begin{split} \label{eq:1}
			u_{ty} - a u_{xy} - b u_{yy} + u_{xxxy} + c (u_{yyyy}- 6v u_{yy}) - 3(u_yu_{xx} + u_{x} u_{xy}) = 0, \\
			u_{yy} - v_{x} = 0,
		\end{split}
	\end{align}
	where $a$, $b$, and $c$ are the arbitrary nonzero constants and $u = u (x, y, t)$ and $v = v (x, y, t)$. We also established that the eBLMP does not pass the
	Painlev\'e test.\hspace{0.1cm}In \cite{101}, Lie's theory
	has been applied to the $(2+1)$-eBLMP system.\hspace{0.1cm}In \cite{14}, the eBLMP equation yields the D\textasciiacute Alembert-type wave and the soliton molecule.
	\subsection{Motivation}
	Having expressed a good amount of information on the eBLMP equation, the motivation behind further studying it comes from the relevance of its solutions in terms of their physical interpretation. Quite a few solutions for the aforementioned problem have already
	been obtained \cite{101,103}. This paper aims to shed light on a few
	new group invariant solutions and the equation's optimal system that have not been studied yet.\\
\vspace{-0.09mm}
 The intriguing fundamental concept behind applying symmetry reductions is the Lie group of transformations' invariance. Due to the one-parameter Lie group's invariance property, we can decrease the independent variables in the given non-linear partial differential equations (NLPDEs) \cite{25}. Consequently, NLPDEs are transformed into ODEs, which give out analytic solutions way more conveniently.
	Numerous aspects of analytical soliton waves seem fascinating, as they have applications in condensed matter physics, non-linear
	waves, plasma physics, non-linear dynamics, oceanography, etc. The novelty of the existence of Lie symmetries is that the invariant functions of the corresponding
	transformation can be determined.\hspace{0.1cm}These invariant functions can be
	used in defining similarity transformations and decreasing the independent variables; in the case of PDEs, the non-linear equation could be written in a more simple
	form.
	\subsection{Framework} The
	theory of Lie is a pioneering method for the investigation of exact
	solutions for nonlinear differential equations. In Section \ref{2}, we apply the similarity transformation to reduce the differential equation.\hspace{0.1cm}Using Lie symmetry, a one-dimensional optimal system for the system \eqref{eq:1} has been presented in Section \ref{3}. Sections \ref{sec:4} and \ref{sec:5} involve the evaluation of symmetry groups, symmetry reductions, and invariant solutions. At last, Section \ref{7} contains our results, and conclusions are drawn through graphs.
	\section{\textbf{Lie Symmetry Analysis of the Non-linear Extended BLMP Equation}} \label{2}
	We consider a one-parameter Lie group with infinitesimal transformations of the independent variables $x$, $y$, and $t$, as well as the dependent variables $u$ and $v$ of the system \eqref{eq:1} as \cite{26}:
	\begin{align}
		\begin{split}
			\check{x} &= x + \varepsilon\xi_{x}(x,y,t,u,v) + O(\varepsilon ^{2}),\\
			\check{y} &= y + \varepsilon\xi_{y}(x,y,t,u,v) + O(\varepsilon ^{2}),\\
			\check{t} &= t + \varepsilon\xi_{t}(x,y,t,u,v) + O(\varepsilon ^{2}),\\
			\check{u} &= u(x,y,t) + \varepsilon\eta_{u}(x,y,t,u,v)+ O(\varepsilon ^{2}),\\
			\check{v} &= v(x,y,t) + \varepsilon\eta_{v}(x,y,t,u,v)+ O(\varepsilon ^{2}).
		\end{split}
	\end{align}
	Then, the Lie symmetry is generated by the following vector field:
	\begin{align}\label{eq:21}
		Z =  \xi_{x}(x,y,t,u,v)\partial_{x} +
		\xi_{y}(x,y,t,u,v)\partial_{y} + \xi_{t}(x,y,t,u,v)\partial_{t}+ \eta_{u}(x,y,t,u,v)\partial_{u} \nonumber \\+ \eta_{v}(x,y,t,u,v)\partial_{v},
	\end{align}
	where $\varepsilon$ is a group parameter, $\xi_{x}$, $\xi_{y}$,
	$\xi_{t}$, $\eta_u$ and $\eta_v$ are the infinitesimals for the variables $x$, $y$, $t$, $u$ and $v$, respectively. For the non-linear eBLMP equation, $Z$ needs to meet the following criteria:
	\begin{align}
		\begin{split}
			Pr^{(4)}Z(\Delta)\lvert_{\Delta_{1}=0} \ = \ 0,
			\\Pr^{(4)}Z(\Delta)\lvert_{\Delta_{2}=0} \ = \ 0,
		\end{split}
	\end{align}
	where $Pr^{(4)}Z $ is the fourth prolongation of $Z$ and
	$\Delta_{1}$, $\Delta_{2}$ are equations from the system
	\eqref{eq:1}.
	Applying the fourth prolongation on \eqref{eq:1}, we get:
	\begin{eqnarray}
		\eta_u^{ty}-a\eta_u^{xy}-b\eta_u^{yy}+\eta_u^{xxxy}+c(\eta_u^{yyyy}-6v\eta_u^{yy}-
		6u_{yy} \eta_v) -3(u_y\eta_u^{xx} \nonumber \\ +u_{xx}\eta_u^{y}+u_x\eta_u^{xy} +u_{xy}\eta_u^{x}) &=0,
		\nonumber \\
		\eta_u^{yy}-\eta_v^{x}&=0, \nonumber
	\end{eqnarray}
	where $\eta_u^{ty}$, $\eta_u^{xy}$, $\eta_u^{yy}$,
	$\eta_u^{xxxy}$, $\eta_u^{yyyy}$ , $\eta_u^{xx}$, $\eta_u^{y}$,
	$\eta_u^{x}$ , $\eta_v$ and $\eta_v^{x}$ are the coefficients of
	$Pr^{(4)}Z$. With the help of these invariant conditions, the total derivative is written as:
	\begin{eqnarray}
		D_i=\dfrac{\partial}{\partial_{x_i}}+u\dfrac{\partial}{\partial_{u}}+u
		_{ij}\frac{\partial}{\partial{u_{ij}}}+ \dots ,\quad i,j = 1,2,3,
		\nonumber
	\end{eqnarray}
	where $x_1$, $x_2$, and $x_3$ are $x$, $y$, and $t$, respectively.
	The obtained equations after some simplifications, are as follows:
	\begin{align}
		\begin{split} \label{eq:22}
			(\xi_{t})_{tt} &= 0, (\eta_{u})_{v} = 0, \\
			(\eta_{u})_{u} &= - \frac{(\xi_{t})_{t}}{3}, \\
			(\eta_{u})_{x} &= - \frac{2a(\xi_{t})_{t}}{9}-
			\frac{(\xi_{x})_{t}}{3}, \\
			(\eta_{u})_{y} &= 0,
			(\xi_{t})_{u} = 0,
			(\xi_{t})_{v} = 0, \\
			(\xi_{t})_{x} &= 0,
			(\xi_{t})_{y} = 0,
			(\xi_{x})_{u} = 0,
			(\xi_{x})_{v} = 0, \\
			(\xi_{x})_{y} &= 0,
			(\xi_{y})_{u} = 0,
			(\xi_{y})_{v} = 0,
			(\xi_{y})_{x} = 0,\\
			(\xi_{x})_{x} &= \frac{(\xi_{t})_{t}}{3},
			(\xi_{y})_{y} = \frac{(\xi_{t})_{t}}{3}, \\
			\eta_{v} &= \frac{(-12c v-2b){((\xi_{t})_{t})-
					3(\xi_{y})_{t}}}{18c}.
		\end{split}
	\end{align}
	The aforementioned set of equations is referred to as the collection of determining equations, which is addressed using MAPLE to get the infinitesimals.
These are the infinitesimals that the aforementioned equations produce:
	\begin{align}
		\xi_{x} &= \frac{1}{3}g_{1}x + F_{1}(t),\nonumber \\
		\xi_{y} &= \frac{1}{3}g_{1}y + F_{2}(t),\nonumber \\
		\xi_{t} &= g_1 t+g_2,\nonumber \\
		\eta_{u} &=\frac{1}{3}xF_{1}^{\prime}(t)+F_3(t)+\frac{1}{9}(-2ax-3u)g_{1},\nonumber \\
		\eta_{v} &= \frac{1}{18}\left[\frac{-3F_{2}^{\prime}+(-
			12cv-2b)g_{1}}{c}\right],\nonumber
	\end{align}
	where $g_1,g_2$ are the arbitrary constants, and $F_{1}(t),F_{2}(t) $
	and $F_{3}(t)$ are the arbitrary functions. We assume the following values:
	$F_{1}(t) = g_{3}$, $F_{2}(t) = g_{4}$, $F_{3}(t) = g_{5} $ where
	$g_3,g_4$, and $g_5$ are the arbitrary constants and substitute these values in the above equations. So, the infinitesimal generators become:
	\begin{align} \label{eq:23}
		\xi_{x} &= \frac{1}{3}g_{1}x + g_{3},\nonumber\\
		\xi_{y} &= \frac{1}{3}g_{1}y + g_{4},\nonumber\\
		\xi_{t} &= g_1 t+g_2,\\
		\eta_{u} &= g_5+\frac{1}{9}\left(-2ax-3u\right)g_{1},\nonumber\\
		\eta_{v} &= \frac{\left(-12cv-2b\right)g_{1}}{18c}. \nonumber
	\end{align}
	Now, the corresponding vector field is rewritten from
	\eqref{eq:21}, keeping in mind information from \eqref{eq:23}, as:
	\begin{align}\label{eq:24}
		Z = \left[ \frac{1}{3}g_{1}x + g_{3}\right] \partial_{x} +\left[ \frac{1}{3}g_{1}y + g_{4}\right] \partial_{y} + (g_1
		t+g_2)\partial_{t} 
		+\left[ g_5+\frac{1}{9}(-2ax-3u)g_{1}\right] \partial_{u}
		+ \\ \left[ \frac{(-12cv-2b)g_{1}}{18c}\right]\partial_{v}. \nonumber
	\end{align}
	Collecting the coefficients of $g_{i}\hspace{0.1cm}(i=1,\dots , 5)$ and
	defining corresponding $v_{i}\hspace{0.1cm}(i=1, \dots, 5)$, the following theorem
	has been deduced.
	\begin{thm}
		The five-dimensional Lie algebra $L_5$ formed by the infinitesimal symmetries for the $(2+1)$-dimensional eBLMP system \eqref{eq:1} is spanned by the following vector fields:
		\begin{align}
			\begin{split} \label{eq:25}
				v_{1} &=\frac{1}{3}x\frac{\partial}{\partial{x}}+\frac{1}{3}y\frac{\partial}{\partial{y}}+t\frac{\partial}{\partial{t}}+\left(\frac{-2ax-3u}{9}\right) \frac{\partial}{\partial{u}}+\left(\frac{-12cv-2b}{18c}\right)\frac{\partial}{\partial{v}},\\
				v_{2}&= \frac{\partial}{\partial{t}},\\
				v_{3} &= \frac{\partial}{\partial{x}},\\v_{4} &= \frac{\partial}{\partial{y}},\\
				v_{5}&= \frac{\partial}{\partial{u}}.
			\end{split}
		\end{align}
	\end{thm}
	Considering the vector fields \eqref{eq:25}, and computing a corresponding one-dimensional optimal system of Lie subalgebras for the eBLMP equation,\hspace{0.1cm}the generators $v_i\hspace{0.1cm}(i = 1,\hspace{0.1cm}2,\hspace{0.1cm}3,\hspace{0.1cm}4,\\ \hspace{0.1cm}5)$ evaluated in \eqref{eq:25} are linearly independent. Consequently, any infinitesimal of system \eqref{eq:1} can be written in the form of a linear combination of $v_i\hspace{0.1cm}(i = 1,\hspace{0.1cm}2,\hspace{0.1cm}3,\hspace{0.1cm}4,\hspace{0.1cm}5)$ as follows:
	\begin{align} \label{eq:26}
		Z = a_{1}v_{1} + a_{2}v_{2} + a_{3}v_{3}+ a_{4}v_{4}+a_{5}v_{5}.
	\end{align}
This Lie algebra, produced via picking linear pairings of $v_1, v_2$, $v_3$, $v_4$, and $v_5$, has a limitless number of sub-algebras. From the author\textasciiacute s work in article \cite{21}, \textquotedblleft if two sub-algebras are similar,
	i.e., connected by a transformation
	from a symmetry group, then their corresponding invariant
	solutions are connected by the same transformation.\textquotedblright
	Consequently, it would be sufficient to select a representative from each category and combine related sub-algebras into a single class. The term "optimal system" refers to the set of these various typical forms.
	\section{\textbf{Building a one-dimensional optimal subalgebra system}} \label{3}
	We are able to organize the invariant solutions of a PDE using an optimal system. The strategy for constructing a one-dimensional optimal system comprises the evaluation of the following:\\
	1) Commutator table\\
	2) Invariants\\
	3) Adjoint transformation matrix \\
	4) A One-dimensional optimal system of subalgebras
\subsection{Commutator table}The commutation relation of a Lie algebra is described thro-ugh a commutator table, whose value is extracted by the definition: [$v_i$,$v_j$]=$v_i$.$v_j$ - $v_j$.$v_i$, where $i, j =1, 2, 3, 4, 5$ (for further information, see \cite{311}).\\
	\textbf{Table 1.} Commutator Table of Lie Algebra for the
	non-linear eBLMP equation. \label{tab:1}
	$$ \begin{tabular}{|l|l|l|l|l|l|l|}
		\hline
		$*$&$v_1$&$v_2$&$v_3$&$v_4$&$v_5$\\\hline
		$v_{1}$&$0$&$-v_{2}$&$-\frac{v_3}{3}+\frac{2a}{9}v_5$&$-
		\frac{v_4}{3}$&$\frac{v_5}{3}$\\\hline
		$v_2$&$v_2$&0&$0$&$0$&0\\\hline
		$v_3$&$\frac{v_3}{3}-\frac{2a}{9}v_5$&$0$&0&0&0\\\hline
		$v_4$&$\frac{v_4}{3}$&$0$&$0$&0&0\\\hline
		$v_5$&$-\frac{v_5}{3}$&$0$&$0$&0&0\\\hline
	\end{tabular}$$
	\subsection{Calculation of invariants}
	
	From the commutator table, we can evaluate the value of $u_i's$ as follows:
	\begin{align*}
		u_1 &= 0,\\
		u_2 &= -e_2b_1+e_1b_2,\\
		u_3 &= -\frac{e_{3}b_{1}}{3}+\frac{e_{1}b_{3}}{3},\\
		u_4 &= -\frac{e_{4}b_{1}}{3}+\frac{e_{1}b_{4}}{3},\\
		u_5 &= \frac{2a}{9}e_{3}b_{1}-\frac{2a}{9}e_{1}b_{3}+\frac{e_{5}b_{1}}{3}-\frac{e_{1}b_{5}}{3}.
	\end{align*}
	For any ${b}_j$, $j$= $1,2,3,4,5$ we require 
	\begin{eqnarray}
		u_{1}\frac{\partial{\phi}}{\partial{e_{1}}}+ u_{2} \frac{\partial{\phi}}{\partial{e_{2}}}+ u_{3} \frac{\partial{\phi}}{\partial{e_{3}}}+e_{4} \frac{\partial{\phi}}{\partial{e_{4}}}+u_{5}\frac{\partial{\phi}}{\partial{e_{5}}}=0. \nonumber
	\end{eqnarray}
	Collecting the coefficients of all $b_j's$ in the above equations,
	an overdetermined system of PDEs of
	$\phi(e_1,e_2,e_3,e_4,e_5)$ is obtained as:
	\begin{align} \label{eq:333}
		\begin{split}
			b_{1}:-e_{2}\frac{\partial{\phi}}{\partial{e_{2}}}+\frac{2a}{9}e_{3}\frac{\partial{\phi}}{\partial{e_{5}}}-\frac{e_{3}}{3}\frac{\partial{\phi}}{\partial{e_{3}}}-\frac{e_{4}}{3}\frac{\partial{\phi}}{\partial e_{4}}+\frac{e_{5}}{3}\frac{\partial{\phi}}{\partial{e_{5}}}=0,\\ 
			b_{2}: e_{1}\frac{\partial{\phi}}{\partial{e_{2}}}=0,\\ 
			b_{3}: \frac{e_{1}}{3}\frac{\partial{\phi}}{\partial{e_{3}}}-\frac{2a}{9}e_{1}\frac{\partial{\phi}}{\partial{e_{5}}}=0,\\
			b_{4}: \frac{e_{1}}{3}\frac{\partial{\phi}}{\partial{e_{4}}}=0,\\
			b_{5}: -\frac{e_{1}}{3}\frac{\partial{\phi}}{\partial{e_{5}}}=0.
		\end{split}
	\end{align}
 The general invariant function of the symmetry algebra $L_{5}$ is in the format: 
 \begin{equation}
    \phi(e_{1},e_{2},e_{3},e_{4},e_{5}) = F(e_{1}), 
 \end{equation} after the PDEs \eqref{eq:333}  were successfully handled.
	\subsection{Calculation of adjoint table and adjoint transformation matrix}This section aims to find the optimal system using the information from the commutator table and adjoint table of the Lie algebra of the non-linear eBLMP equation. Initially, we formulate the adjoint representation of the Lie group.
	\\From the commutator table, we construct the adjoint
	table, using the given series:
	\begin{eqnarray}
		A\textit{d}(exp(\varepsilon v_i)v_j)=v_j-\varepsilon[v_i,v_j]+\dfrac{1}{2}\varepsilon^{2}[v_i,[v_i,v_j]]- \dots \nonumber
	\end{eqnarray}
	Let $\varepsilon_{i}, (i=1,..,5)$ are real-valued constants and
	$g=e^{\varepsilon_{i}x_{i}}$, then with the help of adjoint
	representation table, we obtain
	\begin{align*}
		M_1 =
		\begin{bmatrix}
			1& 0& 0& 0& 0\\
			0& e^{\varepsilon_1}& 0& 0& 0\\
			0& 0&  e^{\frac{\varepsilon_1}{3}}& 0&-\frac{2}{9}a \varepsilon_1\\
			0& 0& 0& e^{\frac{\varepsilon_1}{3}}& 0\\
			0& 0& 0& 0& e^-{\frac{\varepsilon_1}{3}}
		\end{bmatrix},\hspace{1cm}\qquad
		M_2 =
		\begin{bmatrix}
			1& -\varepsilon_2& 0& 0& 0\\
			0& 1& 0& 0& 0\\
			0& 0& 1& 0& 0\\
			0& 0& 0& 1& 0\\
			0& 0& 0& 0& 1
		\end{bmatrix},\\
		M_3 =
		\begin{bmatrix}
			1& 0& \frac{-\varepsilon_3}{3}& 0& \frac{2a}{9}\varepsilon_3\\
			0& 1& 0& 0& 0\\
			0& 0& 1& 0& 0\\
			0& 0& 0& 1& 0\\
			0& 0& 0& 0& 1
		\end{bmatrix},\quad
		M_4 =
		\begin{bmatrix}
			1& 0&  0& -\frac{\varepsilon_4}{3} &0\\
			0& 1& 0& 0& 0\\
			0& 0& 1& 0& 0\\
			0& 0& 0& 1& 0\\
			0& 0& 0& 0& 1
		\end{bmatrix},\quad
		M_5 =
		\begin{bmatrix}
			1& 0&  0& 0& \frac{\varepsilon_5}{3}\\
			0& 1& 0& 0& 0\\
			0& 0& 1& 0& 0\\
			0& 0& 0& 1& 0\\
			0& 0& 0& 0& 1
		\end{bmatrix}.
	\end{align*}
	\textbf{Table 2.} Adjoint Table of Lie Algebra.
	$$\begin{tabular}{|l|l|l|l|l|l|}
		\hline
		$*$&$v_1$&$v_2$&$v_3$&$v_4$&$v_5$\\\hline
		$ v_{1}$&$v_1$&$v_{2}e^{\varepsilon}$&$v_{3}e^{\frac{\varepsilon}{3}}-\frac{2a}{9}v_{5}\varepsilon$&$v_{4}e^{\frac{\varepsilon}{3}}$&$v_{5}e^{\frac{-\varepsilon}{3}}$\\\hline
		$v_2$&$v_{1}-\varepsilon v_{2}$&$v_2$&$v_3$&$v_4$&$v_5$\\\hline
		$v_3$&$v_{1}-\frac{\varepsilon}{3}v_{3}+\frac{2a}{9}\varepsilon v_5$&$v_{2}$&$v_{3}$&$v_{4}$&$v_{5}$\\\hline
		$v_4$&$v_{1}-\frac{\varepsilon}{3}v_{4}$&$v_{2}$&$v_{3}$&$v_{4}$&$v_{5}$\\\hline
		$v_5$&$v_{1}+\frac{\varepsilon}{3}v_{5}$&$v_{2}$&$v_{3}$&$v_{4}$&$v_{5}$\\\hline
	\end{tabular} \label{tab:2}$$
 \subsection{Optimal system of subalgebras in single dimension} \label{sub:3.4}
	Now, to construct the required system, consider $Z =
	\displaystyle{\sum_{i=1}^{5}} a_i v_i$ and $K =
	\displaystyle{\sum_{i=1}^{5}} d_i \Psi_i$ as two elements of Lie
	algebra $L_5$. System (1.1)'s adjoint transformation equation after Z and K has been replaced as follows:
	\begin{equation} \label{eq:34}
		(a_1, a_2, a_3, a_4, a_5) M = (\Psi_1, \Psi_2, \Psi_3, \Psi_4, \Psi_5),
	\end{equation}
	provides the values of $\varepsilon_i\hspace{0.1cm}(i = 1,\hspace{0.1cm}2,\hspace{0.1cm}3,\hspace{0.1cm}4,\hspace{0.1cm}5)$. Here, M is the general adjoint transformation matrix defined as the product of $M_i's$.
	
	\[ \left[ \begin{array}{ccccc}                                              
		a_1& a_2& a_3& a_4& a_5\\
	\end{array} \right]
	\left[ \begin{array}{ccccc}
		1& -\varepsilon_2&  -\frac{\varepsilon_3}{3}& -\frac{\varepsilon_4}{3}& \frac{\varepsilon_5}{3}+\frac{2a}{9}\varepsilon_3\\
		0& e^{\varepsilon_1} & 0& 0& 0\\
		0& 0& e^{\frac{\varepsilon_1}{3}}& 0& -\frac{2a}{9}\varepsilon_1\\
		0& 0& 0& e^{\frac{\varepsilon_1}{3}}&  0\\
		0& 0& 0& 0& e^{\frac{-\varepsilon_1}{3}}
	\end{array} \right] = (\Psi_1, \Psi_2, \Psi_3, \Psi_4, \Psi_5),\]
	or
	\begin{eqnarray} \label{eq:35}
		\Big[a_1, -a_1\varepsilon_2+a_2e^{\varepsilon_1},-a_1\frac{\varepsilon_3}{3}+a_3e^{\frac{\varepsilon_1}{3}}, \ -a_1\frac{\varepsilon_4}{3}+a_4e^\frac{\varepsilon_1}{3}, \nonumber \\ \ a_1\left(\frac{\varepsilon_5}{3}+\frac{2a}{9}\varepsilon_3\right)  +a_3\left(\frac{-2a}{9}\varepsilon_1\right)+a_5\left(e^{\frac{-\varepsilon_1}{3}}\right)  \Big] 
		&= (\Psi_1, \Psi_2, \Psi_3, \Psi_4, \Psi_5).
	\end{eqnarray}
	\textbf{Case 1:}
	Taking $a_1 \not= 0 = l_1 = \{-1,1\}$. From \eqref{eq:34} , assuming $\psi_1 = 1$, $\psi_2 = 0$, $\psi_3=0$, $\psi_4 = 0$, $\psi_5 =0$, and $\varepsilon_1 = 0$, we obtain $\varepsilon_2 = \frac{a_2}{a_1}$, $\varepsilon_3=\frac{3a_3}{a_1}$, $\varepsilon_4=\frac{3a_4}{a_1}$, and $\varepsilon_5=\frac{-(3a_5+2a a_3)}{a_1}$.
	Consequently, $Z = a_1v_1$ .\\
	\textbf{Case 2:} Taking $a_1 = 0$ and $\varepsilon_1 = 0$. 
	Consequently,
	$Z = a_2v_2+a_3v_3+a_4v_4+a_5v_5$.\\
	To summarise, the optimal system is:
	\begin{align} \label{eq:306}
		\begin{split}
			Z = a_1v_1,
			Z = a_2v_2+a_3v_3+a_4v_4+a_5v_5.
		\end{split}
	\end{align}
	\section{\textbf{Symmetry groups of eBLMP equation}} \label{sec:4}
Again, in order to find new analytic solutions to system \eqref{eq:1}, we start with computing one-parameter symmetry groups, \hspace{0.1cm}$ H_i:(x,\hspace{0.1cm}y,\hspace{0.1cm}t,\hspace{0.1cm}u,\hspace{0.1cm}v) \rightarrow{(x^*,\hspace{0.1cm}y^*,\hspace{0.1cm}t^*,\hspace{0.1cm}u^*,\hspace{0.1cm}v^* )}$, produced using $v_i$ $(i = 1,\hspace{0.1cm}2,\hspace{0.1cm}3,\hspace{0.1cm}4,\hspace{0.1cm}5)$, the infinitesimal generators. To accomplish this, we must solve the system of ODEs with initial conditions:
	\begin{align} \label{eq:36}
		\begin{split}
			\frac{dx^*}{d\varepsilon} &= \xi_x(x,y,t,u,v) , x^*_{|\varepsilon=0}=x, \\ \frac{dy^*}{d\varepsilon} &= \xi_y(x,y,t,u,v) , y^*_{|\varepsilon=0} =y, \\
			\frac{dt^*}{d\varepsilon} &=\xi_t(x,y,t,u,v), t^*_{|\varepsilon=0} =t, \\
			\frac{du^*}{d\varepsilon} &= \eta_u(x,y,t,u,v) , u^*_{|\varepsilon=0} =u , \\ 
			\frac{dv^*}{d\varepsilon} &= \eta_v(x,y,t,u,v) , v^*_{|\varepsilon=0} =v. 
		\end{split}
	\end{align}
	
	Then, the one-parameter symmetry groups $H_i$ $(i = 1,\hspace{0.1cm}2,\hspace{0.1cm}3,\hspace{0.1cm}4,\hspace{0.1cm}5)$ are given as follows after solving the system of  ODEs \eqref{eq:36}
	\begin{align} \label{eq:37}
		\begin{split}
			H_1& : (x, y, t, u, v) \rightarrow \left( e^\frac{\varepsilon}{3}
			x, e^\frac{\varepsilon}{3} y, e ^\varepsilon t, -\frac{ax}{3} e^\frac{\varepsilon}{3}+ \left( u+\frac{ax}{3} \right) e^\frac{-\varepsilon}{3},\left( \frac{6cv+b}{6c} \right) e^\frac{-\varepsilon}{3}-\frac{b}{6c}\right),\\
			H_2 &: (x, y, t, u, v) \rightarrow (x, y, t+\varepsilon, u, v), \\
			H_3 &: (x, y, t, u, v) \rightarrow (x+\varepsilon, y, t, u, v),\\
			H_4 &: (x, y, t, u, v) \rightarrow (x, y+\varepsilon, t, u, v),\\
			H_5 &: (x, y, t, u, v) \rightarrow (x, y, t, u+\varepsilon, v).
		\end{split}
	\end{align} \\
   \vspace{-0.09mm}
 The transformed point $exp(\varepsilon Xi(x,\hspace{0.1cm}y,\hspace{0.1cm}t,\hspace{0.1cm}u,\hspace{0.1cm}v)) =(x^*,\hspace{0.1cm}y^*,\hspace{0.1cm}t^*,\hspace{0.1cm}u^*,\hspace{0.1cm}v^*)$ corresponds to the right-sided terms of equations \eqref{eq:37}, where $H_2,H_3,H_4$, and $H_5$ are space translations and $H_1$ is a scale translation. Using the preceding symmetry groups $H_i\hspace{0.1cm} (i =1, 2, 3, 4, 5)$, the relevant new solutions $u_i$ and $v_i$ $(i = 1,2,3,4,5)$ may be obtained as follows \cite{15}: \\
	If $u = f_1(x, y, t), $ $v = f_2(x, y, t)$ is a known solution of \eqref{eq:1}, then by using the above symmetry groups, $H_i\hspace{0.1cm} (i =1, 2, 3, 4, 5)$, the corresponding new solutions $u_i$ and $v_i$ $(i = 1, 2, 3, 4, 5)$ can be obtained as follows \cite{15}:
	\begin{align} \label{eq:38}
		\begin{split}
			u_1 &=-\frac{ax}{3} +\left( f_1(xe^\frac{-\varepsilon}{3},ye^\frac{-\varepsilon}{3},te^{-\varepsilon}) + \frac{ax}{3}e^\frac{-\varepsilon}{3} \right)e^\frac{-\varepsilon}{3},\\
			u_2 &= f_1(x,y,t-\varepsilon),\\
			u_3 &= f_1(x-\varepsilon,y,t),\\
			u_4 &= f_1(x,y-\varepsilon,t),\\
			u_5 &= f_1(x,y,t).\\
			v_1 &= \left(f_2(xe^\frac{-\varepsilon}{3},ye^\frac{-\varepsilon}{3},te^{-\varepsilon})-\frac{b}{6c}\right)e^\frac{-\varepsilon}{3}-\frac{b}{6c},\\
			v_2 &= f_2(x,y,t-\varepsilon),\\
			v_3 &= f_2(x-\varepsilon,y,t),\\
			v_4 &= f_2(x,y-\varepsilon,y,t),\\
			v_5 &= f_2(x,y,t).
		\end{split}
	\end{align}
	
	\section{\textbf{Reductions of symmetry and new solutions}} \label{sec:5}
	For the eBLMP equation, we built a one-dimensional optimal system of Lie subalgebras. Based on the degree of the invariants, we take into consideration the optimal system evaluated in subsection \eqref{eq:306}.
	\subsection{\textbf{Case 1: }} For, $Z = a_{1}v_{1}$,
 retrieving the value of $v\textsubscript{1}$ from \eqref{eq:25},we get
	\begin{eqnarray} \label{eq:82}
		Z =a_1\left[ \frac{1}{3}x\frac{\partial}{\partial{x}}+\frac{1}{3}y\frac{\partial}{\partial{y}}+t\frac{\partial}{\partial{t}}+\left(\frac{-2ax-3u}{9}\right)\frac{\partial}{\partial{u}}+\left( \frac{-12cv-2b}{18c}\right)\frac{\partial}{\partial{v}}\right].
	\end{eqnarray}
	The corresponding auxiliary equation for \eqref{eq:82} is:
	\begin{eqnarray}\label{eq:83}
		\dfrac{dx}{\frac{a_1x}{3}}=\dfrac{dy}{\frac{a_1y}{3}}=\dfrac{dt}{a_1t}=\dfrac{du}{a_1\left(\frac{-2ax-3u}{9}\right)}=\dfrac{dv}{a_1\left(\frac{-12cv-2b}{18c}\right)}.
	\end{eqnarray}
	Integrating \eqref{eq:83}, the following similarity variables and respective similarity functions have been obtained:
 \begin{equation}
	 X =  \frac{x}{y}, T =  \frac{y^3}{t},  f(X,T) = uy +\frac{axy}{3} \text{ and } g(X,T)= \left(v+\frac{b}{6c}\right)y^2.
\end{equation}
	With the proposed dependent variables $f$ and $g$ and independent variables $X$ and $T$, the equations \eqref{eq:1} reduce to the following equation using MAPLE software.
	\begin{multline} \label{eq:84}
 24cf+216cT^3f_{TTT}-9Tf_{T}f_{XX}-9Tf_{X}f_{XT}+XT^2f_{XT}+72cT^2f_{TT}-24cf_{T}+cX^4f_{XXXX}
		-Xf_{XXXX}\\+3ff_{XX}+36gcXTf_{XT}-6gcX^2f_{XX}-24gcXf_{X}+3Tf_{XXXT} -4f_{XXX}
		+81cT^4f_{TTTT} -3T^3f_{TT} \\-2T^2f_T + 6f_X^2 -12cX^3Tf_{XXXT}+16cX^3f_{XXX}-108cXT^3f_{XTTT}
		+ 54cX^2T^2f_{XXTT} -72cX^2Tf_{XXT} \\+ 6Xf_{X}f_{XX}+72cX^2f_{XX}+96cXf_{X}-96cXTf_{XT}-54gcT^2f_{TT}-12gcf=0,
	\end{multline}
	\begin{eqnarray} 
		9T^2f_{TT}-6XTf_{XT}- g_{X}+4Xf_{X}+2f+X^2f_{XX}=0. \nonumber
	\end{eqnarray}
	In order to reduce the system further, we apply the Lie symmetry method and find its infinitesimals:\\
	\begin{eqnarray} \label{eq:86}
		\xi_{X}=\frac{-AX+3B}{3T^1/3},
		\xi_{T}=AT^\frac{2}{3},\eta_{f}=\frac{-BTX+3Af+9CT^2/3}{9T^1/3},\eta_{g}=\frac{36Acg-AT}{54cT^1/3}.
	\end{eqnarray}
	Using \eqref{eq:86}, the auxiliary equations are written as:
	\!
	\begin{eqnarray}\label{eq:87}
		\dfrac{dX}{\frac{-AX+3B}{3T^1/3}}=\dfrac{dT}{A T^\frac{2}{3}}=\dfrac{df}{\frac{-BTX+3Af+9CT^2/3}{9T^1/3}}=\dfrac{dg}{\frac{36Acg-AT}{54cT^1/3}}.
	\end{eqnarray}
	Integrating \eqref{eq:87}, the next set of similarity variables and similarity functions is obtained, which are elaborated in the following cases:\\ 
	\textbf{Case 1.1:} Evaluating the auxiliary equations with $A\not=0$, $B\not=0$, and $C\not=0$, the similarity variable and similarity functions for this particular case are:
	\begin{eqnarray}\label{eq:88}
		s_1 = T^\frac{1}{3}\left( X-\frac{3B}{A} \right), g = \frac{-T}{18c}+T^\frac{2}{3}V(s_1), \\ f = -TX \left( \frac{B}{3A} \right)+\frac{T}{2}\left( \frac{B^2}{A^2}\right)
		+ U(s_1)T^\frac{1}{3}+3T^\frac{2}{3}\left(\frac{C}{A}\right).
	\end{eqnarray}
	Here $U(s_1)$ and $V(s_1)$ are the similarity functions corresponding to the similarity variable $s_1$. On substituting these in \eqref{eq:84}, we get 
	\begin{align}\label{eq:101}
		-&162A^2B^2T^\frac{5}{3}U''(s_1)V(s_1)c + 54A^3BT^\frac{5}{3}U''(s_1)U(s_1) - 18A^2B^2T^\frac{5}{3}V(s_1)c    \nonumber\\
		+& 243B^4T^\frac{5}{3}U''''(s_1) c + 9A^3BT^\frac{5}{3}U'(s_1)  - 9A^3BT^\frac{5}{3}U''''(s_1) - 27A^3CT^\frac{5}{3}U''(s_1)  \nonumber\\
		-& 6A^3CT^\frac{5}{3}+ 6A^3BT^2XU''(s_1)+ A^3BXT^2 
		- 18A^2B^2T^2U''(s_1) - 3A^2B^2T^2 = 0, \nonumber\\
		-&TA^2V'(s_1) + 9TB^2U''(s_1) + TB^2 = 0.
	\end{align}
The system \eqref{eq:101} gives constant solutions. On putting constant values back in the source equation, the final invariant solution comes out as:
\begin{align}
u(x,y,t)= \frac{-ax}{3}-\frac{xy}{t}+\frac{y^2}{2t}+\frac{1}{t^{1/3}}+\frac{3y}{t^{2/3}}, \label{eq:1A}\\
v(x,y,t)= -\frac{y}{18ct}+\frac{R}{t^\frac{2}{3}} - \frac{b}{6c}.\label{eq:1B}
\end{align}
	\textbf{Case 1.2:} Evaluating the auxiliary equations with $A\not=0$, $B=0$, and $C\not=0$, the similarity variable and similarity functions for this particular case are:
	
	\begin{eqnarray}\label{eq:89}
		s_2 = T^\frac{1}{3} X,\hspace{0.2cm}f = U(s_2)T^\frac{1}{3}+3T^\frac{2}{3}\left( \frac{C}{A}\right),\hspace{0.2cm}g = \frac{-T}{18c}+T^\frac{2}{3}V(s_2).	
	\end{eqnarray}
	\\Here $U(s_2)$ and $V(s_2)$ are the similarity functions corresponding to the similarity variable $s_2$. On substituting these in \eqref{eq:84}, we get:
	\begin{eqnarray}
		-9CT^\frac{5}{3}U''(s_2)-2CT^\frac{5}{3}=0, \nonumber
		\\-TV'(s_2)=0,
	\end{eqnarray}
	which gives out the following general solution: \\
	$U = - ((s_2)^2)/9+ M s_2 +N, $\\
	$V = R,$ \\
	where $M,N,$ and $R$ are constants of integration.\hspace{0.1cm}Thus, the invariant solution for the system \eqref{eq:1} is:
 \begin{align}
			u = -\frac{ax}{3}-\frac{x^2}{9t} + \frac{Mx}{t^\frac{2}{3}}+\frac{N}{t^\frac{1}{3}}+\frac{3yC}{At^\frac{2}{3}} ,\label{eq:47a}
			\\
			v = -\frac{y}{18ct}+\frac{R}{t^\frac{2}{3}} - \frac{b}{6c}.\label{eq:47b}
	\end{align}
	\textbf{Case 1.3:} Evaluating the auxiliary equations with $A\not=0$,$B\not=0$, and $C=0$, the similarity variable and similarity functions for this particular case are:
 \begin{eqnarray}\label{eq:90}
		s_3 = T^\frac{1}{3} (X-\frac{3B}{A}),\hspace{0.2cm}f = -T^\frac{2}{3}s_3\left(\frac{B}{3A}\right)+\frac{T}{2}\left(\frac{B^2}{A^2}\right)
		+ U(s_3)T^\frac{1}{3},\hspace{0.2cm}g = \frac{-T}{18c}+T^\frac{2}{3}V(s_3).	\nonumber
	\end{eqnarray}
	\\Here $U(s_3)$ and $V(s_3)$ are the similarity functions corresponding to the similarity variable $s_3$.\\On substituting these in \eqref{eq:84}, we get:
	\begin{align} \label{eq:91}
		\begin{split} 
			& -162A^2B^2T^\frac{5}{3}U''(s_3)V(s_3)c + 54A^3BT^\frac{5}{3}U''(s_3)U(s_3) - 18A^2B^2T^\frac{5}{3}V(s_3)c + 243B^4T^\frac{5}{3}U''''(s_3) c \\
			&+ 9A^3BT^\frac{5}{3}U'(s_3) - 9A^3BT^\frac{5}{3}U''''(s_3) + 6A^3BXU''(s_3)s_3+ A^3Bs_3=0,\\
			&-A^2V'(s_3) + 9B^2U''(s_3) + B^2 = 0.
		\end{split}
	\end{align}
	The system \eqref{eq:91} is a special case of the system \eqref{eq:101}.
	\subsection{\textbf{Case 2:}} For $Z=a_{2}v_{2}+a_{3}v_{3}+a_{4}v_{4}+a_{5}v_{5}$,
	the respective auxiliary equation is 
\begin{eqnarray}\label{eq:873}
		\dfrac{dx}{a_3}=\dfrac{dy}{a_4}=\dfrac{dt}{a_2}=\dfrac{du}{a_5}=\dfrac{dv}{0}.
	\end{eqnarray}
	Equations \eqref{eq:873} after integration give
	$x = X + A_2t$, $y= Y+B_2t$, $u=C_2t+U(X,Y)$, and $v=V(X,Y)$. Here, $X$ and $Y$ are similarity variables, and $U$ and $V$ are similarity functions. Substituting them into  our original system \eqref{eq:1}, we get the following system of PDEs:
	\begin{align}\label{eq:92}
		-(A_2+a)U_{XY}-(B_2+b)U_{YY}+U_{XXXY}+c(U_{YYYY}-6VU_{YY})-3(&U_YU_{XX}+U_{X}U_{XY})=0 ,\nonumber
		\\U_{YY} -& V_X = 0.
	\end{align}
	Here, $A_2$ by $\frac{a_3}{a_2}$,\hspace{0.2cm}$B_2$ by $\frac{a_4}{a_2}$, and $C_2 = \frac{a_5}{a_2}$. The corresponding infinitesimals are evaluated as
	\begin{eqnarray}\label{eq:93}
		\xi_{X}={j_1X+j_2},\hspace{0.2cm}
		\xi_{Y}={j_1Y+j_3},\hspace{0.2cm}\eta_{U}=\frac{((-2A_2-2a)X-3U)j_1+3j_4}{3},\hspace{0.2cm}\eta_{V}=\frac{-j_1(6cV+B_2+b)}{3c}. \nonumber
	\end{eqnarray}
Considering different values of $j_i's$, framing and solving respective cases are as follows: \\ \\
	\textbf{Case 2.1:} Taking $j_1=0$\hspace{0.2cm},$j_2\not=0$,\hspace{0.2cm}$j_3\not=0$ and $j_4\not=0$,
	the infinitesimals become:
	\begin{eqnarray}
		\xi_{X}={j_2},\hspace{0.2cm}
		\xi_{Y}={j_3},\hspace{0.2cm}\eta_{U}={j_4},\hspace{0.2cm}\eta_{V}={0}. \nonumber	
	\end{eqnarray}
	Replacing $ p$ by $\frac{j_2}{j_3}$ and $q$ by $\frac{j_4}{j_3}$ for convenience in $X=pY+P,U=qY+W(P)$, and $V = Z(P),$
	and substituting it in the system \eqref{eq:92},
	we get:
	\begin{align} \label{eq:95}
		[(A_2+a)p-(B_2+b)p^2-3q] W''(P)+(-p+cp^4) W''''+ 2p W'W''-&6cp^2 Z(P) W''(P)=0, \nonumber
		\\ \hspace{-2 cm} p^2 W''(P)-&Z'(P)=0.
	\end{align}
	For this system of ODEs \eqref{eq:95}, a particular solution is:
	\begin{eqnarray}\label{eq:96}
		W =  -\frac{\alpha h_1 P+\gamma h_1 h_2+\alpha h_2}{\gamma h_1},	\nonumber
		\\ Z = p^2\left(\frac{-\alpha}{\gamma}\right) +L.
	\end{eqnarray}
	The terms $\alpha , \beta$, and $\gamma$ are  combinations of constants, whereas $L$, $h_1$, and $h_2$ are  constants of integration.\\
	$\alpha = (A_2+a)p-p^2(B_2+b)-3q-6cp^2 L$,\\
	$\beta = (-p+cp^4)$,\\
	$\gamma = -6p^2+2p$. \\
	The solution for the eBLMP equation \eqref{eq:1} corresponding to the particular solution \eqref{eq:96} is:
 \begin{align}
			u = \left(c-qB_2+\frac{\alpha A_2}{\gamma}+\frac{\alpha p B_2}{\gamma}\right) t-\left(\frac{\alpha}{\gamma}\right) x + \left(q+\frac{\alpha p}{\gamma}\right) y + h_2 + \frac{\alpha h_2}{\gamma h_1},\label{eq:A} \\
			v = p^2\left(-\frac{\alpha}{\gamma}\right)+L.\label{eq:B}
   \end{align}
	\textbf{Case 2.2:} Taking $A=-a$ and $B=-b$, the equation becomes:
	\begin{eqnarray}\label{eq:97}
		U_{XXXY}+c(U_{YYYY}-6VU_{YY})-3(U_YU_{XX}+U_{X}U_{XY})=0,\nonumber \\
		U_{YY} - V_X = 0. \nonumber
	\end{eqnarray}
 A particular solution to this system is
 \begin{align}
 \begin{split}
 U(X,Y)&=-4 \tanh{(2X+3Y+1)}+4,\nonumber\\
 V(X,Y)&=\frac{2[27c\tanh{(2X+3Y+1)^2} -18c+8][\tanh^2{(2X+3Y+1)}-1]}{9c[\tanh{(2X+3Y+1)^2}-1]}.
 \end{split}
 \end{align}
 The back substitution algorithm gives the final solution of \eqref{eq:1}:
 \begin{align} 
 u(x,y,t)&=2t+4-2\tanh{(2x+3y+1)},\label{eq:-2} \\
 v(x,y,t)&= \frac{2[27c\tanh{(2x+3y+1)^2} -18c+8][\tanh^2{(2x+3y+1)}-1]}{9c[\tanh{(2x+3y+1)^2}-1]}\label{eq:-1}.
 \end{align}
\section{Conservation Laws: Multiplier Approach}

Conservation laws in simple mathematics state that there is a certain quantity--let's consider it energy--that does not change in the manifold changes that nature undergoes. It claims that there is a numerical quantity that is constant irrespective of events. It's just a fascinating fact that we may compute a number, and when we repeat the calculation after observing some event, the result remains the same. \\ 
More precisely, the energy conservation law is a theorem associated with quantities that need to be computed and added together \cite{71}.\\
Let $(x,y,t)$ be three independent variables and $(u,v)$ be dependent variables.
\\\\
\textbf{(1)}.~The total derivative operators $D_t$, $D_x$, and $D_y$ are :
\begin{subequations}
	\begin{align}
		D_t&=\frac{\partial}{\partial t}+u_t \frac{\partial}{\partial u}+v_t \frac{\partial}{\partial v}+u_{tt} \frac{\partial}{\partial u_t}+v_{tt}\frac{\partial}{\partial v_t}+u_{tx}\frac{\partial}{\partial u_x}+v_{tx}\frac{\partial}{\partial v_x}+u_{ty}\frac{\partial}{\partial u_y}+v_{ty}\frac{\partial}{\partial v_y},\label{2a} \\
		D_x&=\frac{\partial}{\partial x}+u_x \frac{\partial}{\partial u}+v_x \frac{\partial}{\partial v}+u_{xx} \frac{\partial}{\partial u_x}+v_{xx}\frac{\partial}{\partial v_x}+u_{tx}\frac{\partial}{\partial u_t}+v_{tx}\frac{\partial}{\partial v_t}+u_{xy}\frac{\partial}{\partial u_y}+v_{xy}\frac{\partial}{\partial v_y},\label{2b}\\
		D_y&=\frac{\partial}{\partial y}+u_y \frac{\partial}{\partial u}+v_y \frac{\partial}{\partial v}+u_{yy} \frac{\partial}{\partial u_t}+v_{yy}\frac{\partial}{\partial v_t}+u_{ty}\frac{\partial}{\partial u_x}+v_{ty}\frac{\partial}{\partial v_x}+u_{yx}\frac{\partial}{\partial u_y}+v_{yx}\frac{\partial}{\partial v_y}.\label{2c}
	\end{align}
\end{subequations}
\textbf{(2)}.~
We define the Euler operators of the form :
\begin{subequations}
	\begin{align}
		\frac{\delta}{\delta u}&=\frac{\partial}{\partial u}-D_{t}\frac{\partial}{\partial u_{t}}-D_{x}\frac{\partial}{\partial u_{x}}-D_{y}\frac{\partial}{\partial u_{y}}+D_{t}^2\frac{\partial}{\partial u_{tt}}+D_{x}^2\frac{\partial}{\partial u_{xx}}+D_{y}^2\frac{\partial}{\partial u_{yy}}\dotsc,\label{3a}\\
		\frac{\delta}{\delta v}&=\frac{\partial}{\partial v}-D_{t}\frac{\partial}{\partial v_{t}}-D_{x}\frac{\partial}{\partial v_{x}}-D_{y}\frac{\partial}{\partial v_{y}}+D_{t}^2\frac{\partial}{\partial v_{tt}}+D_{x}^2\frac{\partial}{\partial v_{xx}}+D_{y}^2\frac{\partial}{\partial v_{yy}}\dotsc.\label{3b}
	\end{align}
\end{subequations}
A system of partial differential equations with two independent and two dependent variables of $kth$ order is examined.
\begin{equation}
	\begin{aligned}\label{4}
		H_1(t,x,y,u,v,u_t,u_x,u_y,\dots,v_t,v_x,v_y,\dots)=0,\\
		H_2(t,x,y,u,v,u_t,u_x,u_y,\dots,v_t,v_x,v_y,\dots)=0.
	\end{aligned}
\end{equation}
\\
\textbf{(3)}.~For a vector ${\mathfrak{F}}=({\mathfrak{F}}^{1},{\mathfrak{F}}^{2},{\mathfrak{F}}^{3})$, we have
\begin{equation} \label{5}
	D_{t}{\mathfrak{F}}^{1}+D_{x}{\mathfrak{F}}^{2}+D_{y}{\mathfrak{F}}^{3}=0,
\end{equation}
for all solutions of $(\ref{4})$ is known as the conserved vector of $(\ref{4})$.\\
\textbf{(4)}.~The multipliers $\Lambda_1(x,y,t,u,v)$, and $\Lambda_2(x,y, t,u,v)$ of the system $(\ref{4})$ have the property:
\begin{equation}\label{6}
	D_{t}{\mathfrak{F}}^{1}+D_{x}{\mathfrak{F}}^{2}+D_{y}{\mathfrak{F}}^{3}=\Lambda_1 H_1+\Lambda_2 H_2,
\end{equation}
for some function $u(x,y,t)$, $v(x,y,t)$\cite{15}.\\
\textbf{(5)}.~ 
We get the determining equations for multipliers $\Lambda_1$, and $\Lambda_2$ when we take the derivative of Eq. (\ref{6}) (see\cite{15}):
\begin{subequations}
	\begin{align}
		\frac{\delta}{\delta u}(\Lambda_1 H_1+\Lambda_2 H_2)=0, \label{7a}\\
		\frac{\delta}{\delta v}(\Lambda_1 H_1+\Lambda_2 H_2)=0,\label{7b}
	\end{align}
\end{subequations}
Equations (\ref{7a}) and (\ref{7b}) apply not only to the solutions of system (\ref{4}), but also to the arbitrary functions $u(x,y,t)$, and $v(x,y,t)$. All local conservation laws have multipliers, which can be obtained from equations (\ref{7a}) and (\ref{7b}). Using (\ref{6}) as the determining equation, conserved vectors can then be obtained methodically. In certain instances, however, once the multiplier is known, it is easy to obtain the conserved vectors using simple operations.
\vspace{-0.2cm}
\section{Conservation Laws of the System} \label{sec5}
Our assumed model can be written as
\begin{equation}
	\begin{cases}
		u_{ty}-au_{xy}-bu_{yy}+u_{xxxy}+c(u_{yyyy}-6vu_{yy})-3(u_yu_{xx}+u_xu_{xy})=0, \nonumber \\
		u_{yy}-v_x=0,
	\end{cases}
\end{equation}
The determining equations for multipliers of the form $\Lambda_1$ and $\Lambda_2$ from (\ref{7a}) and (\ref{7b}) are:
\begin{equation}\label{8}
	\frac{\delta}{\delta u}\bigg[\Lambda_1 \big(u_{ty}-au_{xy}-bu_{yy}+u_{xxxy}+c(u_{yyyy}-6vu_{yy})-3u_yu_{xx}-3u_xu_{xy}\big)+
	\Lambda_2\big(u_{yy}-v_x\big)\bigg]=0,
\end{equation}
\begin{equation}\label{9}
	\frac{\delta}{\delta v}\bigg[\Lambda_1 \big(u_{ty}-au_{xy}-bu_{yy}+u_{xxxy}+c(u_{yyyy}-6vu_{yy})-3u_yu_{xx}-3u_xu_{xy}\big)+
	\Lambda_2\big(u_{yy}-v_x\big)\bigg]=0,
\end{equation}
where (\ref{3a}) and (\ref{3b}), respectively, define the basic Euler operators $\frac{\delta}{\delta u}$ and $\frac{\delta}{\delta v}$. Based on the various ways that the derivatives of $u$ and $v$ might be combined, we divided Equations (\ref{8}) and (\ref{9}). This provided us with a set of simplified equations for $\Lambda_1$ and $\Lambda_2$.
\begin{equation}\label{10}
	\begin{aligned}
		& \Lambda_{1x}=0,~\Lambda_{2x}=0,~\Lambda_{1u}=0, \Lambda_{2u}=0,\Lambda_{1v}=0,~\Lambda_{2v}=6\Lambda_{1}c,\\
		&\Lambda_{1yyyy}=\frac{1}{c}\big(6\Lambda_{1yy}cv+\Lambda_{1yy}b-\Lambda_{1ty}-\Lambda_{2yy}\big).
	\end{aligned}
\end{equation}
After solving the above system (\ref{10}), we get
\begin{equation}\label{11}
	\begin{aligned}
		\Lambda_{1}(x,y,t,u,v)&=F_1(yt),\\
		\Lambda_{2}(x,y,t,u,v)&=6F_1(yt)cv+yF_2(t)+F_3(t)+\int{\big(bF_{1y}-cF_{1yyy}-F_{1t}\big)dy}.
	\end{aligned}
\end{equation}
Eqs. (\ref{6}) and (\ref{11}) give us the following conservation laws satisfying Eq. (\ref{5}):
\begin{equation}\label{12}
	\begin{aligned}
		T_t=&\frac{1}{2}u_yF_{1yt}-\frac{1}{2}uF_{1y},\\
		T_x=&\frac{3}{4}uu_xF_{1y}+\frac{3}{4}uu_{xy}F_{1}+\frac{1}{2}auF_{1y}-\frac{9}{4}u_yu_xF_{1}-\frac{1}{2}au_yF_{1}-\frac{1}{4}u_{xx}F_{1y}+\frac{3}{4}u_{xxy}F_{1}\\&-3cv^2F_1-vyF_2-vF_3-bv\int{F_{1y}dy}+cv\int{F_{1yyy}dy}+v\int{F_{1t}dy},\\
		T_y=&yu_{y}F_{2}-uF_2-bu_yF_{1}+cu_yF_{1yy}+u_yF_3-cu_{yy}F_{1y}+cu_{yyy}F_1\\&+\frac{1}{2}uF_{1t}-\frac{3}{4}u^2_xF_{1}-\frac{1}{2}au_{x}F_{1}-\frac{3}{4}uu_{xx}F_{1}+\frac{1}{4}u_{xxx}F_1+\frac{1}{2}u_tF_1\\&+bu_y\int{F_{1y}dy}-cu_y\int{F_{1yyy}dy}-u_y\int{F_{1t}dy}.
	\end{aligned}
\end{equation}
where $F_1=F_1(y,t)$, $F_2=F_2(t)$, and $F_3=F_3(t)$.
So, these are the cases we examine.\\
\textbf{Case A:} Conservation laws corresponding to $$\Lambda_1(x,y,t,u,v)=F(y,t),~~\Lambda_2(x,y,t,u,v)=6cvF(y,t)+\int\big(bF_y-cF_{yyy}-F_t\big)dy,$$ can be written as: 
\begin{equation}
	\begin{aligned}
		\label{13}
		T_t^{(1)}=&\frac{1}{2}u_yF-\frac{1}{2}uF_{y},\\
		T_x^{(1)}=&\frac{3}{4}uu_xF_{y}+\frac{3}{4}uu_{xy}F+\frac{1}{2}auF_{y}-\frac{9}{4}u_yu_xF-\frac{1}{2}au_yF-\frac{1}{4}u_{xx}F_{y}+\frac{3}{4}u_{xxy}F\\&-3cv^2F-bv\int{F_{y}dy}+cv\int{F_{yyy}dy}+v\int{F_{t}dy},\\
		T_y^{(1)}=&cu_{y}F_{yy}-cu_{yy}F_y+cu_{yyy}F-bu_yF+\frac{1}{2}uF_{t}-\frac{3}{4}u^2_xF-\frac{1}{2}au_{x}F-\frac{3}{4}uu_{xx}F\\&+\frac{1}{4}u_{xxx}F+\frac{1}{2}u_tF+bu_y\int{F_{1y}dy}-cu_y\int{F_{1yyy}dy}-u_y\int{F_{1t}dy},
	\end{aligned}
\end{equation}
where $F=F(y,t)$.\\
\textbf{Case B:} Conservation laws corresponding to $$\Lambda_1(x,y,t,u,v)=0,~~\Lambda_2(x,y,t,u,v)=yF(t),$$ can be written as:
\begin{equation}
	\begin{aligned}
		\label{14}
		T_t^{(2)}=&0,\\
		T_x^{(2)}=&-yvF(t),\\
		T_y^{(2)}=&-uF(t)+yu_yF(t).
	\end{aligned}
\end{equation}
\textbf{Case C:} Conservation laws corresponding to $$\Lambda_1(x,y,t,u,v)=0,~~\Lambda_2(x,y,t,u,v)=F(t),$$ can be written as:
\begin{equation}
	\begin{aligned}
		\label{15}
		T_t^{(3)}=&0,\\
		T_x^{(3)}=&-vF(t),\\
		T_y^{(3)}=&u_yF(t).
	\end{aligned}
\end{equation}
We observed that we can construct infinitely many conserved vectors from all three cases by choosing the specific value of the involved function. If we choose $F(t)=\sin{t}$ for Cases 2 \& 3, then we can write the specific case of the form. \\
\textbf{Subcase (i):} Conservation laws corresponding to $$\Lambda_1(x,y,t,u,v)=0,~~\Lambda_2(x,y,t,u,v)=y\sin{t},$$ can be written as:
\begin{equation}
	\begin{aligned}
		\label{16}
		T_t^{(2)}=&0,\\
		T_x^{(2)}=&-yv\sin{t},\\
		T_y^{(2)}=&-u\sin{t}+yu_y\sin{t}.
	\end{aligned}
\end{equation}
\textbf{Subcase (ii):} Conservation laws corresponding to $$\Lambda_1(x,y,t,u,v)=0,~~\Lambda_2(x,y,t,u,v)=\sin{t},$$ can be written as:
\begin{equation}
	\begin{aligned}
		\label{17}
		T_t^{(3)}=&0,\\
		T_x^{(3)}=&-v\sin{t},\\
		T_y^{(3)}=&u_y\sin{t}.
	\end{aligned}
\end{equation}
	\section{Figures and Conclusions}\label{7}
\begin{figure}[ht]
			\begin{subfigure}
					[3D figure of u]{\includegraphics[width=40mm]{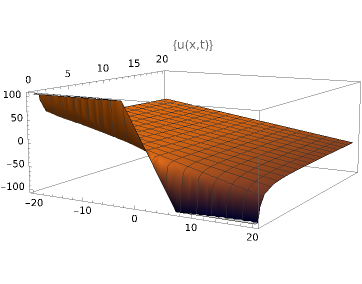}} \label{subfig:1a}
				\end{subfigure}
		\quad
			\begin{subfigure}
					[3D figure of u]{\includegraphics[width=43mm]{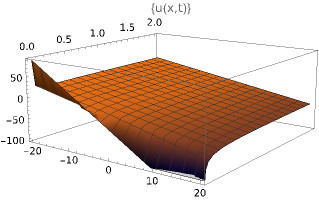}}  \label{subfig:1b}
				\end{subfigure}
		\quad
			\begin{subfigure}
					[3D figure of u] {\includegraphics[width=43mm]{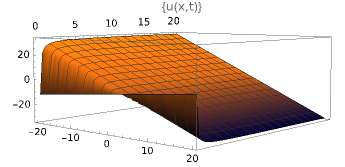}}\label{subfig:1c} \quad
				\end{subfigure} \caption{Graphical analysis of solution in \eqref{eq:1A}} \label{fig:1}
\end{figure}
\begin{figure}
			\begin{subfigure}
					[3D plot of u at t=0.001]{\includegraphics[scale=0.40]{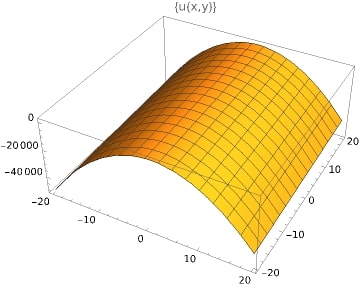}} 
					\label{subfig:2a}
				\end{subfigure}
		\quad
			\begin{subfigure}
					[Contour plot of u at t=0.001]{\includegraphics[width=30mm]{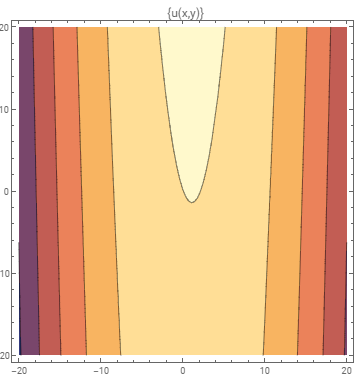}}  \label{subfig:2b}
				\end{subfigure}
		\quad
			\begin{subfigure}[2D plot of u at t=0.001]{\includegraphics[width=35mm]{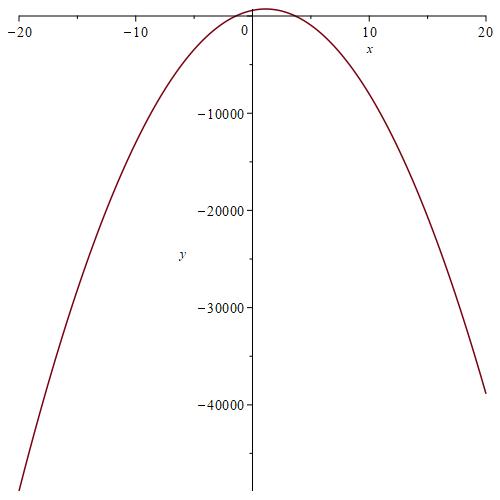} \label{subfig:2c}} 
				\end{subfigure}
		\qquad
			\begin{subfigure}
					[3D plot of u at y=1]{\includegraphics[width=40mm]{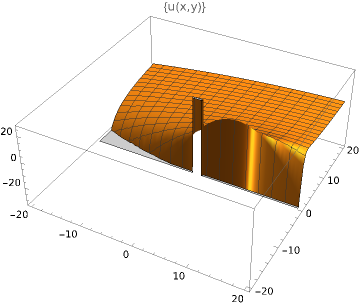}} \label{subfig:2d}
				\end{subfigure}
		\qquad
			\begin{subfigure}
					[Contour plot of u at y=1]{\includegraphics[width=30mm]{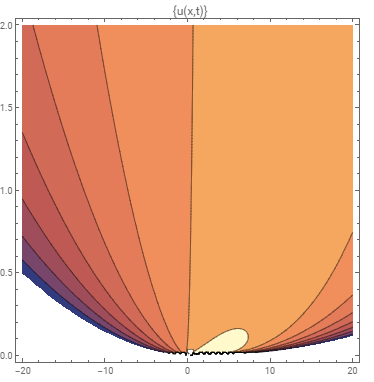}} \label{subfig:2e}
				\end{subfigure}
			\qquad
			\begin{subfigure}
					[2D plot of u at y=1] {\includegraphics[width=40mm]{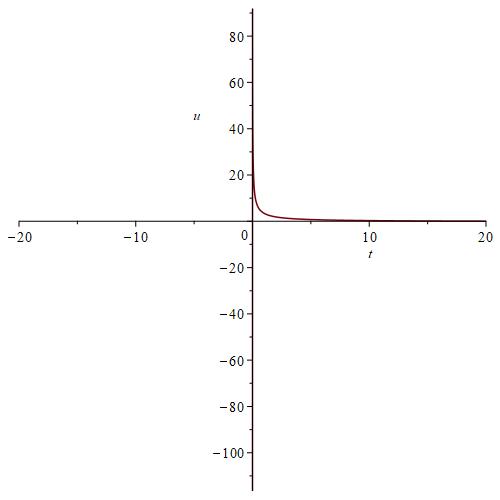}} \label{subfig:2f}
				\end{subfigure}
    \caption {Solution profiles for equations \eqref{eq:47a}} \label{fig:2}
 \end{figure}
 \begin{figure*}[ht]
			\begin{subfigure}
					[3D plot of v]{\includegraphics[width=51mm]{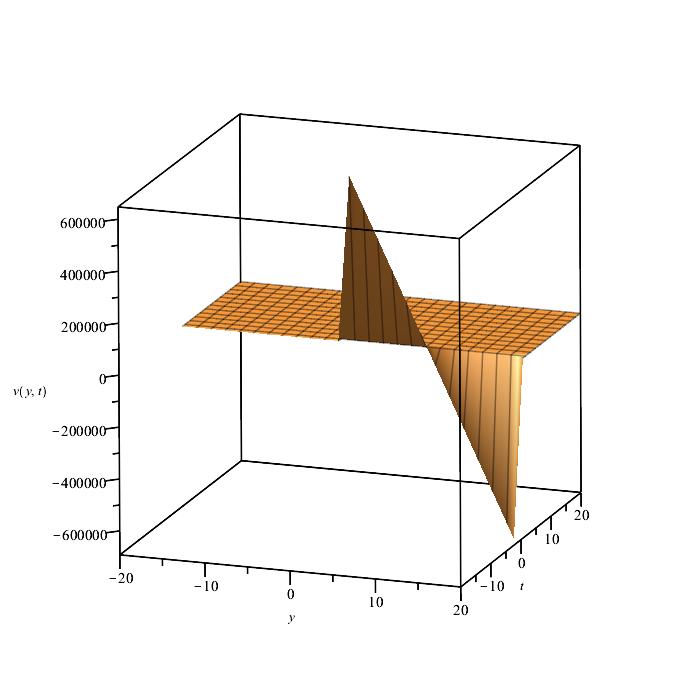}} \label{subfig:3a}
				\end{subfigure}
			\qquad
			\begin{subfigure}
					[Contour plot of v]{\includegraphics[width=39mm]{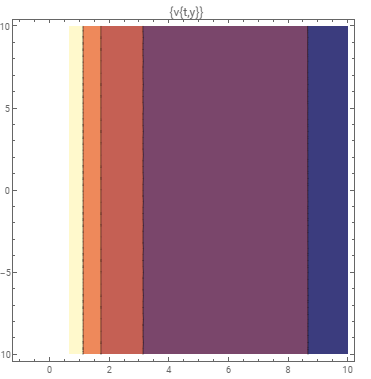}} \label{subfig:3b}
				\end{subfigure}
			\qquad
			\begin{subfigure}
					[2D plot of v]{\includegraphics[width=31mm]{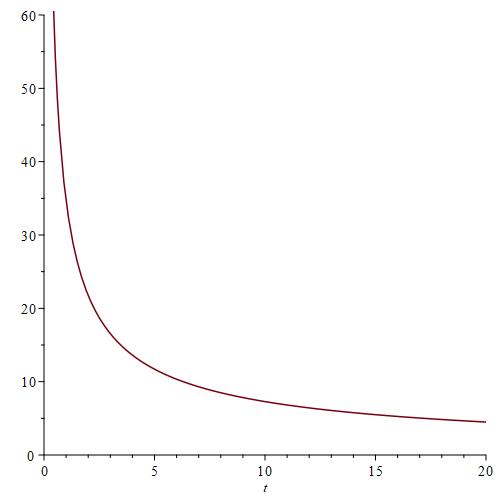}} \label{subfig:3c}
				\end{subfigure}
			\qquad
			\begin{subfigure}
					[3D plot of v]{\includegraphics[width=45mm]{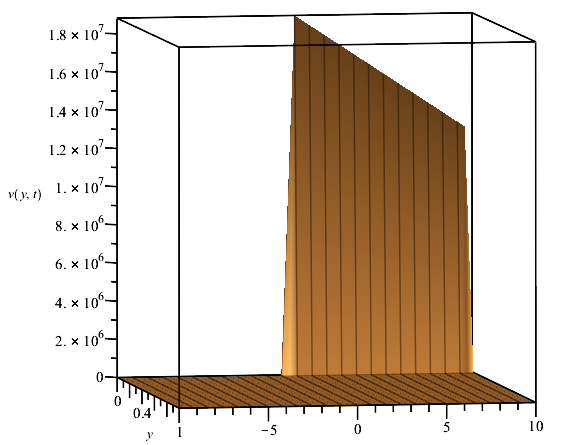}} \label{subfig:3d}
				\end{subfigure}
			\qquad
			\begin{subfigure}
					[Contour plot of v]{\includegraphics[width=38mm]{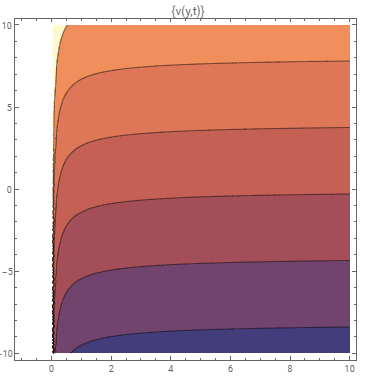}} \label{subfig:3e}
				\end{subfigure}
			\qquad
		\begin{subfigure}
					[2D plot of v]{\includegraphics[width=35mm]{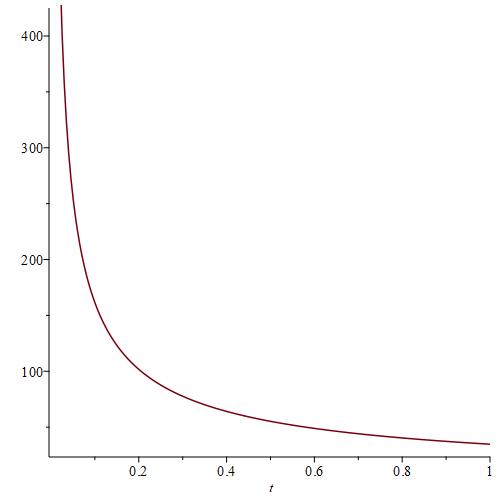}} \label{subfig:3f}
					\caption {Solution profiles 
							for equations \eqref{eq:47b}} \label{fig:3}
				\end{subfigure}
\end{figure*}
	\begin{figure}[hb]
			\centering
			\begin{subfigure}
					[3D plot of u]{\includegraphics[width=45mm]{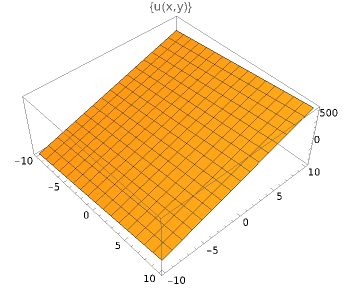}} \label{subfig:4a} \quad
				\end{subfigure}
            \begin{subfigure}
					[3D plot of v at t=1]{\includegraphics[width=45mm]{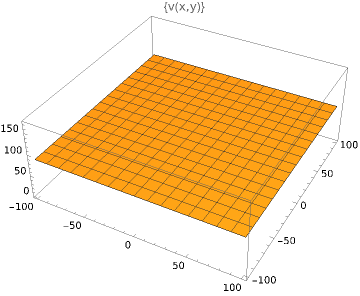}}\label{subfig:4b}
				\end{subfigure}
    \begin{subfigure}
					[3D plot of u at t=1]{\includegraphics[width=45mm]{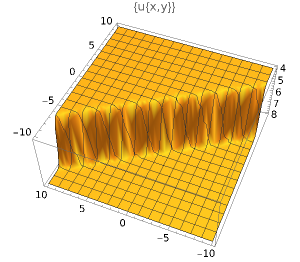}} \label{subfig:4c}
				\end{subfigure}
    \caption {Solution profiles of \eqref{eq:A} and \eqref{eq:B} \label{fig:4}}
    \end{figure}
The Lie-based symmetry paradigm has been applied to the analysis of the eBLMP system. The methodology involves determining a one-parameter Lie group of infinitesimal transformations. We have used MAPLE to obtain infinitesimals that form the foundation of the five-dimensional Lie algebra $L_5$. Additionally, we have presented the commutator table in Section \ref{tab:1} and the adjoint table in Section \ref{tab:2}. The classification of all subalgebras via a one-dimensional optimal system has been elaborated as different cases in subsection \ref{sub:3.4}. To evaluate further, the system of non-linear PDEs \eqref{eq:1} is transformed into systems of ODEs. The systems of ODEs have been solved as cases. To reduce some systems, the theory of Lie is used once more. The reduced system of ODEs, after back substitution, gives invariant solutions of \eqref{eq:1}.

Fig.\ref{fig:1} depicts the solution \eqref{eq:1A} for different values of $y$. The portraits have curved features since shallow water waves have dispersive and non-linear effects.\\
The freedom that arbitrary constants give in solutions provides different graphs. Giving particular values to constants in \eqref{eq:47a} as $M = 2.5, N = 2, C = 2$ and $A= 7$, the graph plotted in fig. \ref{fig:2} turns out to be a parabola. Figs. \ref{fig:2}(a)--(c) describe the solution $u$ given in \eqref{eq:47a} graphically. Fig. \ref{fig:2}(a) shows a parabolic profile, whereas in Fig. \ref{fig:2}(b) a contour diagram and in \ref{fig:2}(c) a two-dimensional figure for $u$ at t = 0.001 are presented. For $y=1$, the graph evolves midway into a parabola; the overall profile seems similar to a warped parabolic profile with a dip in the middle. This profile gives a not localised and time-dependent solution. The contour plot and 2D plot are depicted in Fig. \ref{fig:2}(e) and Fig. \ref{fig:2}(f), respectively. The solutions $u$ and $v$ both have singularities at $t=0$, which can be seen in Figs. \ref{fig:2}(d) and \ref{fig:3}(a). Solutions in \eqref{eq:1B} and \eqref{eq:47b} are the same and are depicted in Figs. \ref{fig:3}. Fig. \ref{fig:3}(a) particularly shows up a double soliton wave profile. The Figs. \ref{fig:3}(a)--(c) show \eqref{eq:47b} for the range $y \in [0,1]$, while Figs. \ref{fig:3}(d)--(f) for $y\in[-20,20].$ \\
The invariant solution \eqref{eq:-2} shows the evolutionary behavior in Fig. \ref{fig:4}(a), which concludes that $u$ varies linearly with time. However, the function $v$ in the solution \eqref{eq:-1} is a constant function; hence, its profile is that of a plane, as shown in Fig. \ref{fig:4}(b). The Fig. \ref{fig:4}(c), at the value $t=1$, holds a stationary portrait, dips, and then continues onto a stationary portrait. This figure suggests the presence of a stationary point as well. This problem can be studied further using the graphs, solutions, and conserved vectors obtained in this article.\\
To summarise, the system \eqref{eq:1} provides three sets of invariant solutions,[\eqref{eq:47a}, \eqref{eq:47b}][\eqref{eq:-2},\\ \eqref{eq:-1}] and [\eqref{eq:A},\eqref{eq:B}].
The one-dimensional optimal system is:
\begin{align} 
	\begin{split}
		Z = a_1v_1, \nonumber
		Z = a_2v_2+a_3v_3+a_4v_4+a_5v_5. \nonumber
	\end{split}
\end{align}
  \vspace{-0.09mm}
The conserved vectors have been evaluated and finalized in \eqref{13}--\eqref{17}.\\
  \vspace{-0.09mm}
  This work establishes the extension of the work done by  Andronikos Paliathanasis in \cite{101}. The additional research here includes distinct symmetries, new solutions, and conserved vectors. This will advance the application of the Lie symmetry paradigm in partial differential equations, especially since conservation laws play a crucial part in both the mathematical and physical arenas. The future scope of this work is to find the killing form for this system. The methodology used in this article can be used to evaluate conservation laws for new problems. The integration of machine learning approaches with optimal systems of PDEs has promising prospects for adaptive control, optimization, and decision-making within dynamic contexts. Synergies across these domains may be investigated in future studies to create more intelligent and adaptable systems.\\
\textbf{Acknowledgement}\\
 The first author feels extremely grateful to “the Ministry of Education, the Government of India," for sponsoring this research work.
 The corresponding author acknowledges the financial support provided by the funding agency CSIR, New Delhi, India, with sanction letter number 25/0327/23/EMR-II dated 03/07/2023.\\
\textbf{Statement of disclosure}\\
 There is no disagreement or dispute regarding this work, among the researchers.


\begin{thebibliography}{30}
\bibitem{1}B. Fritzsche (1999), \textit{Sophus Lie, A sketch of his life and work}, Journal of Lie Theory, 9, pp.1-38.
\bibitem{2} R.K. Gazizov and N.H. Ibragimov (1998), \textit{Lie symmetry analysis of differential equations in finance}, Nonlinear Dynamics, 17, pp.387-407.
\bibitem{3} E. Momoniat, D.P. Mason and F.M. Mahomed (2001), \textit{Non-linear diffusion of an axisymmetric thin liquid drop: group-invariant solution and conservation law}, International Journal of Non-linear Mechanics, 36(6), pp.879-885.
\bibitem{4} H. Liu, J. Li and Q. Zhang (2009), \textit{Lie symmetry analysis and exact explicit solutions for general Burgers’ equation}, Journal of Computational and Applied Mathematics, 228(1), pp.1-9.
\bibitem{30} Malik, S., Almusawa, H., Kumar, S., Wazwaz, A.M. and Osman, M.S., (2021), \textit{A (2+1)-dimensional Kadomtsev–Petviashvili equation with competing dispersion effect: Painlevé analysis, dynamical behavior and invariant solutions}, Results in Physics, 23, pp.104043.
\bibitem{301} Taghizadeh, A., Sarkardeh, H. and Jabbari, E., (2023), \textit{Numerical study on effect of reservoir vortex on velocity distribution profile in pipe}, The European Physical Journal Plus, 138(6), pp.521.
\bibitem{302} Miah, M.M., Iqbal, M.A. and Osman, M.S., (2023), \textit{A study on stochastic longitudinal wave equation in a magneto-electro-elastic annular bar to find the analytical solutions}, Communications in Theoretical Physics, 75(8), pp.085008.
\bibitem{303} Qureshi, S., Akanbi, M.A., Shaikh, A.A., Wusu, A.S., Ogunlaran, O.M., Mahmoud, W. and Osman, M.S., (2023), \textit{A new adaptive nonlinear numerical method for singular and stiff differential problems}, Alexandria Engineering Journal, 74, pp.585-597.
\bibitem{31} Rahman, R.U., Qousini, M.M.M., Alshehri, A., Eldin, S.M., El-Rashidy, K. and Osman, M.S., (2023), \textit{Evaluation of the performance of fractional evolution equations based on fractional operators and sensitivity assessment}, Results in Physics, 49, pp.106537.
\bibitem{32} Djennadi, S., Shawagfeh, N., Osman, M.S., Gómez-Aguilar, J.F. and Arqub, O.A., (2021), \textit{The Tikhonov regularization method for the inverse source problem of time fractional heat equation in the view of ABC-fractional technique}, Physica Scripta, 96(9), pp.094006.
\bibitem{33} Tasnim, F., Akbar, M.A. and Osman, M.S., (2023), \textit{The extended direct algebraic method for extracting analytical solitons solutions to the cubic nonlinear Schrödinger equation involving beta derivatives in space and time}, Fractal and Fractional, 7(6), pp.426.
\bibitem{42} Akinyemi, L., Houwe, A., Abbagari, S., Wazwaz, A.M., Alshehri, H.M. and Osman, M.S., (2023), \textit{Effects of the higher-order dispersion on solitary waves and modulation instability in a monomode fiber}, Optik, 288, pp.171202.
\bibitem{43} Kumar, S., Niwas, M., Osman, M.S. and Abdou, M.A., (2021), \textit{Abundant different types of exact soliton solution to the $(4+1)$-dimensional Fokas and $(2+1)$-dimensional breaking soliton equations}, Communications in Theoretical Physics, 73(10), pp.105007.
\bibitem{44} Tripathy, A., Sahoo, S., Rezazadeh, H., Izgi, Z.P. and Osman, M.S., (2023), \textit{Dynamics of damped and undamped wave natures in ferromagnetic materials}, Optik, 281, pp.170817.
\bibitem{45} Ismael, H.F., Bulut, H., Park, C. and Osman, M.S., (2020), \textit{M-lump, N-soliton solutions, and the collision phenomena for the $(2+1)$-dimensional Date-Jimbo-Kashiwara-Miwa equation}, Results in Physics, 19, pp.103329.
\bibitem{46} Ismael, H.F., Sulaiman, T.A., Nabi, H.R., Mahmoud, W. and Osman, M.S., (2023), \textit{Geometrical patterns of time variable Kadomtsev–Petviashvili (I) equation that models dynamics of waves in thin films with high surface tension}, Nonlinear Dynamics, 111(10), pp.9457-9466.
  \bibitem{51} G. Bluman, A.F. Cheviakov and S.C. Anco, \textit{Applications of Symmetry Methods to Partial Differential Equations}, Springer, New York, 168, 2010.
\bibitem {15} P. Olver, \textit{Applications of Lie Groups to Differential Equations}, Springer-Verlag Inc., New York, 107,1986.
\bibitem{8}A. Paliathanasis and M. Tsamparlis (2018), \textit{Lie symmetries for systems of evolution equations}, Journal of Geometry and Physics, 124, pp.165-169.
\bibitem{9} C.R. Gilson, J.J.C. Nimmo and R. Willox (1993), \textit{A (2+ 1)-dimensional generalization of the AKNS shallow water wave equation}. Physics Letters A, 180(4-5), pp.337-345.
\bibitem{17} M. Kumar and A.K. Tiwari (2018), \textit{Soliton solutions of BLMP equation by Lie symmetry approach}, Computers \& Mathematics with Applications, 75(4), pp.1434-1442.
\bibitem{18} L.V. Ovsiannikov,\textit{Group Analysis of Differential Equations}, Academic Press, New York , 2014.
\bibitem{20} J. Patera, P. Winternitz and H. Zassenhaus (1975), \textit{Continuous subgroups of the fundamental groups of physics. I. General method and the Poincaré group}, Journal of Mathematical Physics, 16(8), pp.1597-1614.
\bibitem{21} J. Patera, R.T. Sharp, P. Winternitz and H. Zassenhaus (1976), \textit{Invariants of real low dimension Lie algebras}, Journal of Mathematical Physics, 17(6), pp.986-994.
\bibitem{22} X. Hu, Y. Li and Y. Chen (2015), \textit{A direct algorithm of one-dimensional optimal system for the group invariant solutions}, Journal of Mathematical Physics, 56(5), pp.053504.
\bibitem{23} Tiwari, A. and Arora, R. (2022), \textit{Lie symmetry analysis, optimal system and exact solutions of a new (2+ 1)-dimensional KdV equation,} Modern Physics Letters B, 36(12), p.2250056.
\bibitem{24} Tiwari, A., Sharma, K. and Arora, R. (2021), \textit{Lie symmetry analysis, optimal system, and new exact solutions of a (3+ 1) dimensional nonlinear evolution equation,} Nonlinear Engineering, 10(1), pp.132-145.
\bibitem{11125} Sharma, A.K., Yadav, S. and Arora, R. (2023), \textit{Invariance analysis, optimal system, and group invariant solutions of (3+ 1)‐dimensional non‐linear MA‐FAN equation,} Mathematical Methods in the Applied Sciences, 46(17), pp.17883-17909.
\bibitem{101}A. Paliathanasis (2022), \textit{Lie symmetry
analysis for a $(2+1)$ extended Boiti-Leon-Manna-Pempinelli equation}, Quaestiones Mathematicae, 46, pp.1-8.
\bibitem{102}H.D. Guo and T.C. Xia (2020), \textit{Lump and Lump-Kink soliton solutions of an extended Boiti-Leon-Manna-Pempinelli equation}, International Journal of Nonlinear Sciences and Numerical Simulation, 21(3-4), pp.371-377. 
\bibitem{103} M. Li, W. Tan and H. Dai (2022), \textit{Integrability tests and some new soliton solutions of an extended potential Boiti-Leon-Manna-Pempinelli equation}, Journal of Applied Mathematics and Physics, 10(10), pp.2895-2905.
\bibitem{111} M. Boiti, J.P. Leon, M. Manna and F. Pempinelli (1986), \textit{On the spectral transform of a Korteweg-de Vries equation in two spatial dimensions}, Inverse problems, 2(3), pp.271.
\bibitem{112} L. Delisle and M. Mosaddeghi (2013), \textit{Classical and SUSY solutions of the Boiti–Leon–Manna–Pempinelli equation}, Journal of Physics A: Mathematical and Theoretical, 46(11), pp.115203.
\bibitem{113} X. Deng, H. Chen and Z. Xu (2014), \textit{Diversity soliton solutions for the (2+ 1)-Dimensional Boiti-Leon-Manna-Pempinelli Equation}, Journal of Mathematics Research, 6(4), pp.85.
\bibitem{114} Z.H. Fu and J.G. Liu (2017), \textit{Exact periodic cross-kink wave solutions for the (2+ 1)-dimensional Boiti-Leon-Manna-Pempinelli equation}, Indian Journal of Pure `I\&' Applied Physics (IJPAP), 55(2), pp.163-167.
\bibitem{117} L. Luo (2011), \textit{New exact solutions and Bäcklund transformation for Boiti-Leon-Manna-Pempinelli equation}, Physics Letters A, 375(7), pp.1059-1063.
\bibitem{121} Y. Li and D. Li (2012), \textit{New exact solutions for the $(2+1)$-dimensional Boiti-Leon-Manna-Pempinelli equation}, Applied Mathematical Sciences (Ruse), 6, pp.579-587.
\bibitem{311} N.K. Ibragimov and N.K. Ibragimov, \textit{Elementary Lie group analysis and ordinary differential equations}, Vol. 197, New York: Wiley, 1999.
\bibitem{14} B. Ren (2021), \textit{Dynamics of a D’Alembert wave and a soliton molecule for an extended BLMP equation}, Communications in Theoretical Physics, 73(3), pp.035003.
\bibitem{25} D. Singh, S. Yadav and R. Arora (2022), \textit{A (2+ 1)-dimensional modified dispersive water-wave (MDWW) system: Lie symmetry analysis, optimal system and invariant solutions}, Communications in Nonlinear Science and Numerical Simulation, 115, pp.106786.
\bibitem{26} M. Devi, S. Yadav and R. Arora (2021),\textit{Optimal system, invariance analysis of fourth-order nonlinear ablowitz-Kaup-Newell-Segur water wave dynamical equation using Lie symmetry approach}, Applied Mathematics and Computation, 404, pp.126230.   
 \bibitem{71} R.P. Feynman, R.B. Leighton and M. Sands, \textit{Six Easy Pieces: Essentials of physics explained by its most brilliant teacher}, 4th ed., Basic Books, 2011. 
\end{thebibliography}
\end{document}